\documentclass[twocolumn]{aastex6}

\usepackage{graphicx}
\usepackage{array}
\usepackage{amssymb}
\usepackage{natbib}
\usepackage{float}
\usepackage{color}
\usepackage[utf8]{inputenc}  
\usepackage{amsmath}  
\usepackage{comment}
\usepackage{footnote}
\usepackage{longtable}
\usepackage[caption2]{ccaption}

\newcommand{\degree}{\ensuremath{^\circ}}

\begin{document}


\title{Flux and polarization variability of OJ 287 during early 2016 outburst}


\author{Suvendu Rakshit \altaffilmark{1,2}, C.S. Stalin \altaffilmark{1}, S. Muneer \altaffilmark{1}, S. Neha \altaffilmark{3, 4} and Vaidehi S. Paliya \altaffilmark{1,5}}


\altaffiltext{1}{Indian Institute of Astrophysics, Block II, Koramangala, Bangalore-560034, India}
\altaffiltext{2}{suvenduat@gmail.com}
\altaffiltext{3}{Aryabhatta Research Institute of Observational Sciences (ARIES), 263002, Nainital, India}
\altaffiltext{4}{Pt. Ravishankar Shukla University, 492010, Raipur, India}
\altaffiltext{5}{Department of Physics and Astronomy, Clemson University, Kinard Lab of Physics, Clemson,
SC 29634-0978, USA}

\begin{abstract}
The gamma-ray blazar OJ 287 was in a high activity state during December 2015 
- February 2016. Coinciding with this high brightness state, we observed this 
source for photometry on 40 nights in $R$-band and for polarimetry on 9 epochs 
in $UBVRI$ bands. During the period of our observations, the source brightness 
varied between $13.20 \pm 0.04$ to $14.98 \pm 0.04$ mag and the degree of 
polarization ($P$) fluctuated between $6.0 \pm 0.3$\% and $28.3 \pm 0.8$\% in 
$R$-band. Focusing on intra-night optical variability (INOV), we find a duty 
cycle of about 71\% using $\chi^2$-statistics, similar to that known for 
blazars. From INOV data, the shortest variability time scale is estimated to 
be $142 \pm 38$ min yielding a lower limit of the observed Doppler factor 
$\delta_0 = 1.17$, the magnetic field strength $B \le 3.8$ G and the size of 
the emitting region $R_s < 2.28 \times 10^{14}$ cm. On inter-night timescales, 
a significant anti-correlation between $R$-band flux and $P$ is found. The 
observed $P$ at $U$-band is generally larger than that observed at longer 
wavelength bands suggesting a wavelength dependent polarization. Using 
$V$-band photometric and polarimetric data from Steward Observatory obtained 
during our monitoring period we find a varied correlation between $P$ and 
$V$-band brightness. While an anticorrelation is seen between $P$ and 
$V$-band mag at sometimes, no correlation is seen at other times, thereby, 
suggesting the presence of more than one short-lived shock components in the 
jet of OJ 287.

\end{abstract}


\section{Introduction}
OJ 287 is a well-known BL Lac object that shows featureless continuum spectrum.
It has been extensively studied for optical flux variability
\citep{1970ApL.....6..201B,1971ApL.....9..151A,1978ApJS...38..267O}. The long term optical light curve  shows a well-defined 
11.65 years of periodicity between large outbursts \citep{1988ApJ...325..628S}. Several models have been 
proposed to explain the periodicity in outburst such as the binary black hole 
model with the primary having an accretion disk 
\citep{1988ApJ...325..628S,1996A&A...315L..13S}, 
quasi-periodic oscillations in an accretion disc \citep{1999MNRAS.303..309I}, 
and a binary black hole  without relativistic precession 
\citep{1997ApJ...478..527K,1998MNRAS.293L..13V,2000ApJ...531..744V}. 
Among them, a precessing binary black hole in which the secondary black hole 
affects the accretion disk of the primary is more favorable than others as 
it predicts more accurately the timing of the major 
outburst \citep{1997ApJ...484..180S,2012MmSAI..83..219V}.
OJ 287 has also been studied for polarization variability 
\citep{1977IzKry..56...39S,1991AJ....101.2017S,1992A&A...254L..33S,2000ApJ...531..744V,2000A&AS..146..141P,2002A&A...381..408E,2009MNRAS.397.1893V}. 
\citet{1991AJ....101.2017S}, based on the observations carried over 
six nights, found  
anticorrelation between flux and 
polarization variations which they explained as a result of highly rotating 
plasma 
inside a relativistic jet. However, \citet{2009MNRAS.397.1893V} did not 
find any clear correlation between flux and polarization. From long term (year time scale) photopolarimetric observations, \citet{2002A&A...381..408E} noticed 
rapid continuous rotation of the position angle of about 4.92 degrees/day in 
clockwise direction suggesting a helical magnetic field jet structure. OJ 287 is also known to show variability and flares at GeV $\gamma$-ray energy \citep{2009ATel.2256....1C,2011ATel.3680....1E,2011MNRAS.412.1389N,2011ApJ...726L..13A}.    

OJ 287 was predicted to have a major outburst in 2015 by 
\cite{2011ApJ...742...22V}. In line with the prediction, many episodes of 
flaring behavior
were noted since December 2015. \citet{2015ATel.8372....1S} and \citet{2016ApJ...819L..37V} 
reported a strong optical flare 
on 05 December 2015,
wherein they found an increase in brightness of about 1.5 mag.  
The WEBT/GASP project 
\citep{2015ATel.8374....1L} reported that the
source reached maximum brightness in the $J$-band on 04 December 2015.
During the same
period, enhanced brightness was also reported by the SMARTS
monitoring program \citep{2015ATel.8382....1M} and also independently by 
\cite{2015ATel.8378....1V}. In the X-ray band too, Swift/XRT observations \citep{2015ATel.8395....1W,2015ATel.8401....1C,2016ApJ...819L..37V} found the source in a high brightness level 
on 05 December 2015. The source was again detected in a flaring activity on 
05 February 2016 
\citep{2016ATel.8667....1Z}. 

We have been monitoring OJ 287 repeatedly for photometric and polarimetric
variations since January 2016 \citep{2016ATel.8697....1P,2016ATel.8806....1M}. 
Here, we present our new $R$-band photometric observations obtained 
during 40 nights from 07 January 2016 to 11 April 2016 including 21 nights of 
intra-night optical variability (INOV) as well as $UBVRI$ polarimetry including the ones already reported by us in \citet{2016ATel.8697....1P} and \citet{2016ATel.8806....1M},  
and $R$-band intra-night polarization variability (INPV) on 3 nights.  
The main motivation behind this monitoring
is to understand (i) the INOV nature of the source in its recent flaring state and (ii) the relation between total flux and polarization characteristics of the source. The paper is organized as follows. In Section \ref{sec:observation} 
we present our observations and analysis, the results of our 
monitoring are reported in Section \ref{sec:result}, followed 
by the discussion in Section \ref{sec:discussion}. We summarize our results in Section \ref{sec:conclusion}. We adopt a cosmology $H_0=70 \, \mathrm{km \, s^{-1}\, Mpc^{-1}}$ and $q_0=0$.

\section{Observation and data reduction}\label{sec:observation}
\subsection{Photometry}
Photometric observations in $R$-band were carried out with a 1k $\times$ 
1k CCD attached to the 0.75-m
telescope at 
the Vainu Bappu 
Observatory (VBO) in Kavalur, India.  
The CCD has a pixel size of 24 microns, image scale of $0.48^{\prime \prime}$/pixel, gain of 1.01 e$^-$ ADU$^{-1}$ and readout noise of 11.51 e$^-$.
Due to weather 
constraints, on some nights we were able to get only few points but on
21 nights we obtained more than 20 frames which allowed us to study  
INOV of the source.  
The source was suitably placed in the CCD so 
as to get at least three comparison stars given in \citet{1996A&AS..116..403F}. 
The log of the photometric observations is given in Table \ref{Table:log}.
The images were analyzed 
using standard procedures in IRAF\footnote{IRAF is 
by the Association of Universities for Research in Astronomy, Inc. under 
cooperative agreement with the National Science Foundation.}. 
To get the optimum 
aperture for aperture photometry we followed the procedure described in 
\citet{2004JApA...25....1S}.

\subsection{Polarimetry}
Polarimetric observations were carried out on a total of 9 nights, of which on 
6 nights single epoch multi-band $UBVRI$ observations were performed and on 
three nights continuous monitoring was done in $R$-band. For 
polarimetric observations, two telescopes were used, one the 104 cm telescope, 
located at VBO and the other the 104 cm Sampurnand telescope located at the Aryabhatta Research Institute for Observational Sciences (ARIES), 
Nainital. At the telescope in VBO, 
a three-band, double beam photo-polarimeter  
was used, the details of which can be found in 
\citet{2015arXiv150504244S}. 
We used diaphragm of 20$^{\prime \prime}$ diameter for the observations. In addition to the $UBVRI$ bands, we also obtained polarimetric measurements in the light integrated in the $V-I$ spectral
region; we refer to this band as $R'$.  At ARIES,
the ARIES Imaging 
Polarimeter (AIMPOL, \citealt{2007MNRAS.378..881M}) was used. A detailed 
description of AIMPOL and the techniques of polarization measurements may be found in \citep{1998A&AS..128..369R,2004BASI...32..159R,2016A&A...588A..45N}. 
All polarimetric data are presented in Table \ref{Table:pol-data}.

\section{Results}\label{sec:result}
\subsection{Intra-night optical variability (INOV)}\label{sec:INOV}
\begin{figure*}
\centering
\resizebox{18cm}{7.5cm}{\includegraphics{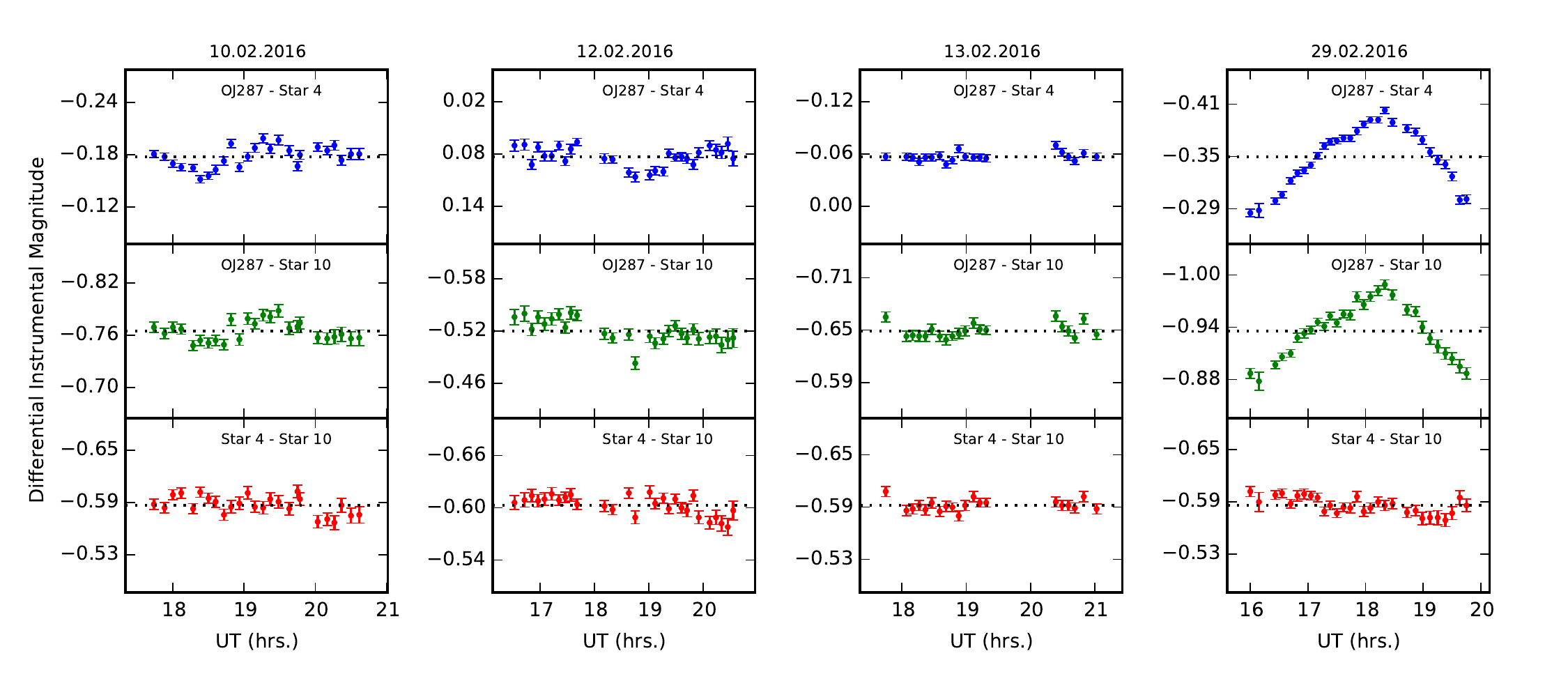}}
\caption{Intra-night DLCs for OJ 287. From top to bottom the DLCs are for OJ 287 - star 4, OJ 287 - star 10 and star 4 - star 10. The dates of observations are written on the top of each panel. The dotted black lines indicate the mean of the DLC. Stars 4 and 10 are those given by \citet{1996A&AS..116..403F}. Only 4 intra-night DLCs are shown here and the other 17 DLCs are available in the electronic format only.}\label{Fig:DLC_1}
\end{figure*}

To study INOV, we restricted to only observations carried out for a minimum of 
about two hours so as to ensure the availability of a sufficient number of
photometric points to characterize INOV. 
DLCs of the OJ 287 was 
generated relative to two comparison stars present on the same CCD frame
as described in Section \ref{sec:observation}.  We note that the chosen 
optimum aperture for photometry on each night is often close to the median FWHM and the host galaxy has negligible effects in
our  photometry \citep{2000AJ....119.1534C}. Some DLCs are shown in 
Figure \ref{Fig:DLC_1}.   
In the star-star DLCs (with the comparison stars having 
similar brightness to OJ 287)  at certain epochs, deviant points are noticed 
due to non-photometric sky conditions. Such data points are identified if they are greater than 2$\sigma$, where $\sigma$ is the standard deviation of the 
star-star DLCs. The number of such deviant points that are removed amounts to 
maximum of two data points each in less than half a dozen of observing nights.
To ascertain the variability nature of OJ 207 on any given night,
we have employed three criteria outlined below.

One method is based on the parameter $C$ given by \citet{1997AJ....114..565J}. 
It is defined as the ratio of the standard deviation of the 
source-comparison star ($\sigma_\mathrm{s}$) and the comparison 
stars ($\sigma$) DLCs and is given as
$C=\sigma_\mathrm{s}/{\sigma}$. As the DLCs of OJ 287 were generated 
relative to two comparison
stars, we obtained two values of $C$. The source is considered variable 
only when both the values of $C \ge 2.576$ 
\citep[see, ][]{2013MNRAS.428.2450P}. 

As an alternative to the widely used $C$-statistics, \citet{2010AJ....139.1269D} 
proposed the $F$-statistics.  
It is defined as the ratio of the variance of 
source-comparison star ($\sigma^{2}_\mathrm{s}$) and the 
comparison stars ($\sigma^{2}$) DLCs and is given by
$F=\sigma^{2}_\mathrm{s}/{\sigma^{2}}$.
To find the variability on any given night using $F$ value, we compared both 
the $F$ values (relative to the two comparison stars)  with the critical 
$F$ value, $F^{\alpha}_{\nu}$, where $\alpha$ is 
the significance level and $\nu$ is the degrees of freedom ($\nu=N_p -1$ where 
$N_p$ is the number of data points in the DLC). Following 
\cite{2013MNRAS.428.2450P}, we used $\alpha=0.01$, which corresponds to a 
probability $p \ge 99 \%$. The source is considered to be variable only if both 
the $F$ values are greater than $F^{\alpha}_{\nu}$. 

We also used $\chi^2$-statistics 
\citep{1976AJ.....81..919K} to characterize INOV. According to this, if 
the $\chi^2$ value of a DLC exceeds the critical value, 
$\chi^2_{\alpha, \nu}$, with significance $\alpha=0.01$, then the source is 
considered variable. $\chi^2$-statistics is defined by
\begin{equation}
\chi^2=\sum_{i=1}^{n} \frac{(D_i - <D>)^2}{\epsilon^2_i}
\end{equation}
Here $\epsilon_i$ is the error of the measurement $D_i$, and $<D>$ is defined as
\begin{equation}
<D>=\frac{\sum_{i=1}^{n}\epsilon^{-2}_i D_i}{\sum_{i=1}^{n} \epsilon^{-2}_i}
\end{equation}   

We calculated the amplitude of variability 
($\Psi$, \citealt{1999A&AS..135..477R}) from the DLCs as  
$ \Psi =100 \sqrt{(D_{\mathrm{max}} - D_{\mathrm{min}})^2 - 2\sigma^2} \, 
\mathrm{\%}$. Here, $D_{\mathrm{max}}$ and  $D_{\mathrm{min}}$ are the  
maximum 
and the minimum in the DLC of OJ 287 relative to the comparison
stars and  $\sigma^2$ is the variance of the star-star DLC.
Thus, corresponding to the two DLCs of the source with respect to the 
two comparison stars, we have two values of $\Psi$ on each night. 
The results of the $C$, $F$ and $\chi^2$-statistics and $\Psi$ for all the 
21 DLCs are given in Table \ref{Table:INOV_prop}. 
 
We also estimated the duty cycle ($DC$) of INOV of OJ 287 using the definition 
of \citet{1999A&AS..135..477R},
\begin{equation}
DC = 100 \frac{\sum_{i=1}^{n} N_i (1/\Delta t_i)}{\sum_{i=1}^{n} (1/\Delta t_i)} \, \mathrm{per\,cent}, 
\end{equation}
where $\Delta t_i= \Delta t_{i, \mathrm{obs}} (1+z)^{-1}$ is the duration of 
the monitoring session of the source on the $i^{\mathrm{th}}$ night after cosmological redshift 
($z$) correction. If INOV is detected then $N_i=1$, otherwise $N_i=0$. We find 
an INOV $DC=30$ \% when variability was characterized using $C$-statistics.
However, using $F$-statistics the $DC$ increased to 45\%, and further increased to 71 \% considering $\chi^2$-statistics. This enhanced 
$DC$ is similar to what is known for blazars
\citep{2004JApA...25....1S}. 

We calculated the minimum variability time scale in our INOV data  as 
$\tau = dt/{\mathrm{ln}(F_1/F_2)}$
following the definition given by \citet{1974ApJ...193...43B}.
Here, $dt$ is the time difference between any two flux measurements
$F_1$ and $F_2$. From our observed DLCs we calculated all possible 
time differences $\tau_{ij}$ for all allowable pairs of observations
for which $|F_i - F_j| > \sigma_{F_i} + \sigma_{F_j}$. From the ensemble 
of $\tau_{ij}$ values, the minimum time scale is obtained as
$\tau_{\mathrm{var}} = \mathrm{min}\left[\tau_{i,j} \right]$. 
Here, $i$ runs from 1 to $n-1$, and $j$ runs from $i+1$ to $n$, where $n$ is 
the total
number of data points. The uncertainties in the $\tau_{i,j}$ values
are determined by propagating the errors in the flux measurements \citep{1969drea.book.....B}. Using this method on all the DLCs where 
INOV is detected we find a minimum $\tau_{\mathrm{var}}$ of 
$142 \pm 38$ min in the observations done on 07 April 2016.

 \subsection{Long-term optical variability (LTOV)}\label{sec:LTOV} 
\begin{figure*}
\centering
\resizebox{15cm}{7.0cm}{\includegraphics{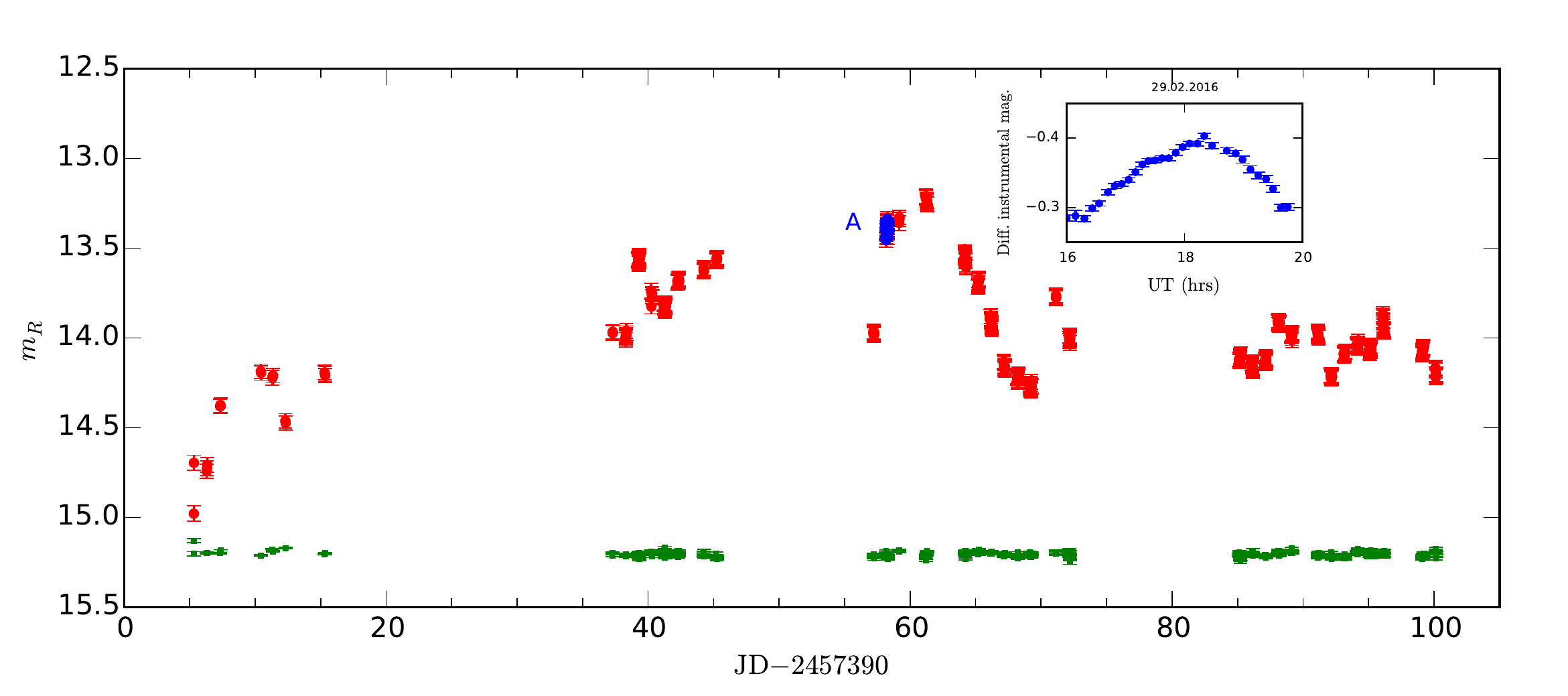}}
\caption{Long term variation of $R$-band magnitude. The bunch of points in the plots are due to the intra-night monitoring, one such bunch, denoted by ``A'' includes the DLC of 29 February 2016, shown in the right inset plot. The DLC of star 4 and star 10 is shown at the bottom (square) after shifting by 15.8 mag.} 
\label{Fig:long_lc}
\end{figure*}

 The time span of our monitoring program is large enough to search for LTOV. The LTOV light curve of OJ 287 from 07 January 2016 to 11 April 2016 is shown in 
 Figure \ref{Fig:long_lc}. The magnitude of OJ 278 was calibrated 
 using the three standard stars as mentioned in section \ref{sec:INOV}.
 Figure \ref{Fig:long_lc} shows OJ 287 is variable on day like time scale. During our monitoring program a change of about 2 
 mag was found within a few days.

 \subsection{Polarization variability}
\begin{figure*}
\centering
\resizebox{15cm}{6.5cm}{\includegraphics{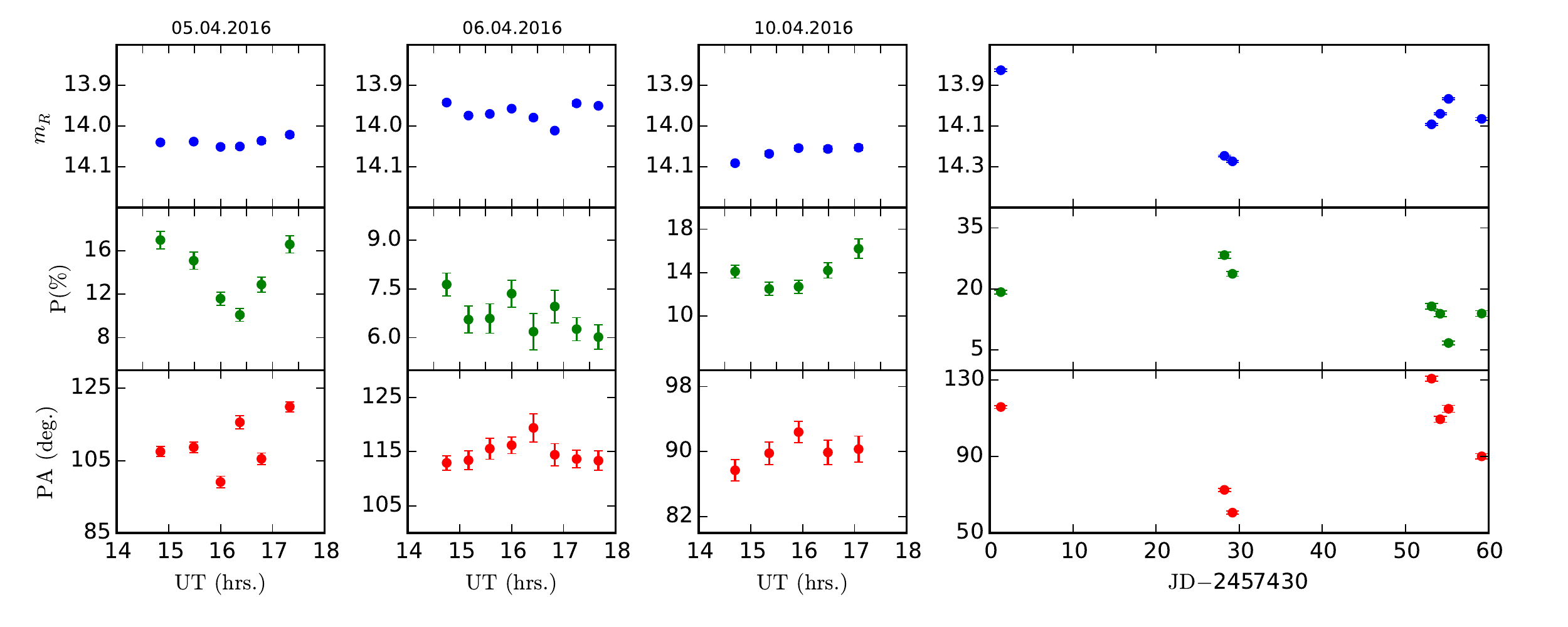}}
\caption{Intra-night polarization variability of OJ 287. Shown from top to bottom are the variation of $R$-band magnitude ($m_R$), the degree of polarization (\%) and position angle (deg.) as a function of UT (hours.). The dates are shown in the top of each panel. In the right most panel is shown the LTPV of OJ 287.}\label{Fig:intra_pol}. 
\end{figure*}

 Intra-night polarization variability (INPV) of OJ 287 has been studied 
 earlier by \citet{2009MNRAS.397.1893V} who found about 16 \% polarization.  
 On the nights of 05, 06 and 10 April 2016, we have sufficient data points
 in $R$-band to characterize the INPV of OJ 287. The polarization properties displayed by the source on those three nights are plotted in the 1st three panels of Figure 
 \ref{Fig:intra_pol}.  We also have in total 7 epochs of $R$-band polarization measurements 
 between February and April 2016. These observations
 are shown in the last panel of Figure \ref{Fig:intra_pol}. 
 When more than one measurements are available on any particular night, we have 
 taken their average value to study Long-term polarization variability (LTPV). From this Figure, it is clear that the source 
 has shown INPV as well as LTPV.

 We also studied the correlation between different observed quantities and the results are shown in Figure  \ref{Fig:intra_pol_rmag}. The solid lines in these Figures are
 the linear least-squares fit to the data.
 A correlation between brightness and $P$ is found on  
05 April 2016. Less INPV was observed on 06 April 2016 with $P$ changing by only 1.5 \% while 
 $PA$ changed by about 7 degrees. On 10 April 2016, the source becomes fainter 
 by about 0.2 mag than its brightness on 06 April 2016, however, $P$ 
 increased by 7\% and $PA$ decreased by about 20 degrees. 

 On the LTPV, we find a clear anticorrelation between source brightness and $P$.  However, $PA$ is found to be correlated with brightness. We also find a negative correlation between $PA$ and $P$. Similar results have been found by \citet{1991AJ....101.2017S}. The statistics of the correlation analysis between the photometric and polarimetric properties of OJ 287 are shown in Table \ref{Table:Pol_prop}.

 To characterize INPV in our data we used the $\chi^2$-statistics (see section \ref{sec:INOV}). We considered the source as variable in polarization if $\chi^2$ value exceeds the critical value $\chi^2_{\alpha, \nu}$ with significance $\alpha=0.01$.  
 The fractional variability (FV) index of the source is defined by
 \begin{equation}
 FV=\frac{A_{\mathrm{max}} - A_{\mathrm{min}}}{A_{\mathrm{max}}+A_{\mathrm{min}}}.
 \end{equation} 
 Here, $A_{\mathrm{max}}$ and $A_{\mathrm{min}}$ are the maximum and minimum 
amplitude of variations in both  $P$ and $PA$. The results of INPV are given 
in Table \ref{Table:INPV_prop}. The $DC$ of INPV is found to be 81 \%, which is 
similar to the $DC$ of about 77 \% found by \citet{2005A&A...442...97A} for 
radio selected BL Lacs for which OJ 287 belongs to.
 
\begin{figure*}
\centering
\resizebox{11cm}{5.0cm}{\includegraphics{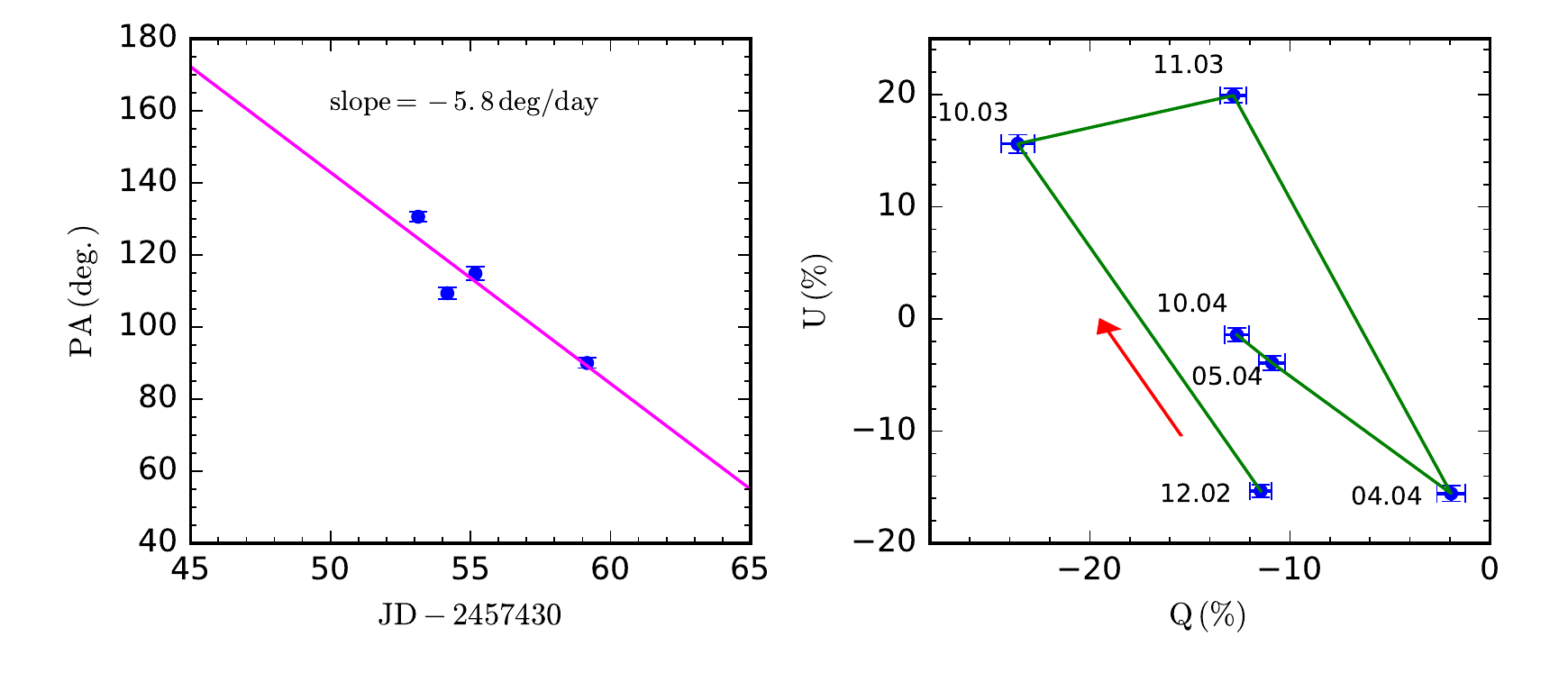}}
\caption{Left: Rotation rate of polarization position angle. A straight line has been fit to the April 2016 polarization data. The estimated slope is $-5.8$ degree/day. Right: Plot of the equatorial Stokes Q and U parameters in the Q-U plane in $R$-band. The arrow indicates the direction of rotation of the plane of polarization. The labels indicate the date of observations. }\label{Fig:rate}
\end{figure*}

Figure \ref{Fig:rate} shows the long term variation of $PA$ observed in the month of April 2016. Linear least squares fit to the data gives a rotation rate of 5.8 
 degree/day. This rate is close to the rate of  
 4.92 degree/day, found by \citet{2002A&A...381..408E}. The observed Q and U parameters are plotted in Figure \ref{Fig:rate}. The average values of Q and 
 U are $<\mathrm{Q}> = -12.2 \pm 0.2$ \% and $<\mathrm{U}> = -0.1 \pm 0.2 $ \% respectively. This deviates
 from the origin implying the presence of a stable polarized component 
\citep[see, ][]{1985ApJ...290..627J}.
 
 \subsection{Wavelength dependent polarization (WDP)}
\begin{figure*}
\centering
\resizebox{9cm}{11cm}{\includegraphics{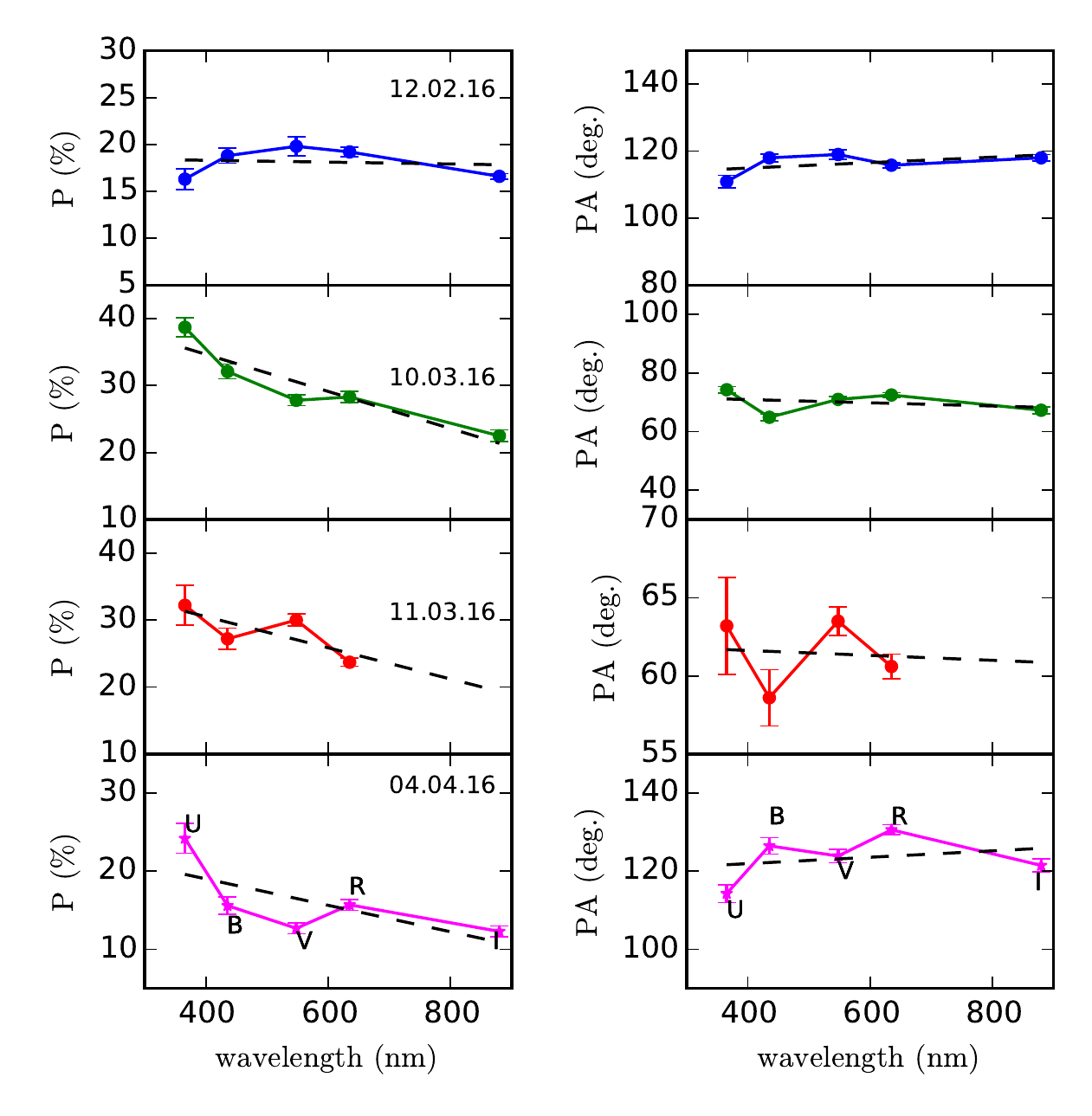}}
\caption{Wavelength dependent polarization. Degree of polarization (left panel) and polarization position angle (right panel) are plotted as a function of wavelength for different dates of observations. The dashed line shows the linear fit to the data. Filter names are marked in the lower panels.}\label{Fig:pol_freq}
\end{figure*}

 On few epochs, we have
 polarization observations in $UBVRI$ bands.
 The multi-band polarization
 variations are shown in Figure \ref{Fig:pol_freq}. 
 During 
 all the epochs except that on 12 February 2016, $P$ in $U$-band is larger
 than the other bands and thus OJ 287 showed WDP. The variation
 of $P$ as a function of wavelength for different nights is shown in Figure \ref{Fig:pol_freq}.
From the Figure, it is clear that on some epochs, both
 $P$ and $PA$ decreases with wavelength, although the anticorrelation of $P$ and wavelength is stronger than $PA$ with wavelength.

 \section{Discussion}\label{sec:discussion}
 
 OJ 287 has shown remarkable optical flux and polarization variations
 during its recent bright state in December 2015 $-$ April 2016. We find
 the source to show large amplitude and high duty cycle of INOV.
 A large amplitude flare (0.12 mag) with 
a slow rise and fast declining pattern 
was found on 29 February 2016. 
Though the exact mechanisms for the cause of INOV are not known, the  observations
 reported here can be used to put constraints on the physical characteristics of 
 the 
 source. From our INOV observations, we find the shortest time scale of 
 variability, $\tau_{\mathrm{var}}$ of $142 \pm 38 $ min on 07 April 2016
that  sets an upper limit on the size of the emission region, 
$R_s  < 19.5 \times 10^{14} (\delta/10)$ cm, where $\delta$ is the Doppler factor.
 
We estimated the observed Doppler factor, $\delta_o$, based on 
relativistic beaming model \citep{1980PASP...92..127M,
2007ApJ...664L..71A,2007PASJ...59.1061X}. Following 
\citet{1979PASP...91..589B}, the observed monochromatic flux ($f_R$) is 
calculated from the apparent $R$-band magnitude ($m_R$) of OJ 287 
(see Figure \ref{Fig:long_lc}) as   
 $f_R=3.08 \times 10^{-23} 10^{-0.4 m_R}\, \mathrm{W \, m^{-2}\, Hz^{-1}}$.
 The observed source-frame monochromatic luminosity ($L_{\nu_s}$) at the 
frequency $\nu_s$  (considering $\nu_s$ as the $V$-band frequency) 
is calculated from $f_R$ using  
 \begin{equation}
 L_{\nu_V}= 4\pi D_L f_R \left[\frac{\lambda_R}{\lambda_V (1+z)} \right]^{\alpha} (1+z)^{-1}
 \end{equation} 
 where the luminosity distance $D_L=(cz/H_0)^2 (1+z/2)^2$, $\lambda_R$ and $\lambda_V$ are the effective central wavelengths of $R$ and $V$ band respectively, and $\alpha$ is the spectral index. We used $\alpha=1.62$, which is the average spectral index found by \citet{2002A&A...381..408E}. Though blazars show 
spectral variations, the value of $\alpha$ used here is similar to the 
value found for OJ 287 and other blazars from power law fits to broad band
optical data \citep{2014ApJ...789..135W}.
 
 Following \citet{1994ApJS...95....1E}, we estimated the observed bolometric 
luminosity as
 $L_B = 13.2 \nu_{V} L_{\nu_V}$
 where $\nu_V$ is the $V$-band frequency.
 Considering the fact that any strong outburst having energy $\Delta L=|L_i-L_j|$ must occur on timescale larger than $t_{\mathrm{min}}=\tau_{\mathrm{var}}/(1+z) << t_{\mathrm{cross}}$ (light crossing time of $R_s$), the inferred 
efficiency of accretion, $\eta_o$ can be calculated as
$\eta_o \ge 5 \times 10^{-43} \Delta L/t_{\mathrm{min}}$ 
\citep{1979xras.proc..381F}.
For our observed $\tau_{\mathrm{var}}=142$ min, we find 
$t_{\mathrm{min}}=108$ min, during which the bolometric luminosity has 
changed 
by $\Delta L=4.33 \times 10^{45} \, \mathrm{erg\,s^{-1}}$ corresponding to $\eta_o=0.33$. In the 
case of disk accretion, a rapidly rotating black hole has an intrinsic value 
of accretion efficiency ($\eta_i$) less than about 0.3 
\citep{1986S&T....71..579F}. As our calculated value is greater than 0.1, 
the observed INOV is due to  relativistic beaming. 
 
 The $\delta_o$ can be computed from the relationships of $\Delta L (o) = 
\delta^{3+\alpha} \Delta L (i)$ and $t_{\mathrm{min}} 
(o)= \delta^{-1} t_{\mathrm{min}} (i)$ 
\citep{1986S&T....71..579F,2002PASJ...54..159Z}, and using 
$\eta_o \ge 5 \times 10^{-43} \frac{\Delta L (o)}{t_{\mathrm{min}} (o)}$ 
and $\eta_i \ge 5 \times 10^{-43} \frac{\Delta L (i)}{t_{\mathrm{min}} (i)}$ 
we find
 \begin{equation}
 \delta_{o} \ge \left(\frac{\eta_o}{\eta_i} \right)^{\frac{1}{4+\alpha}},
 \end{equation}
 where `$o$' and `$i$' refer to the observed and intrinsic values. Since the value of $\eta_i$ can be between 0.007 (nuclear reaction) to 0.32 (maximum accretion), we used $\eta_i=0.05$ (a geometric mean value) in the above equation and found $\delta_o \ge 1.17$. Using this $\delta_o$ and observed $\tau_{\mathrm{var}}$ 
of 142 min, we found $R_s < 2.28 \times 10^{14} $ cm. 
 
 Considering that the variable emission seen in OJ 287 is due to synchrotron 
processes, and requiring that $\tau_{\mathrm{var}}$ to be shorter than the synchrotron lifetime of the relativistic electrons in the observer frame \citep{2008ApJ...672...40H}, the magnetic field ($B$) can be estimated as 
 \begin{equation}
 t_{\mathrm{syn}} \propto  4.75 \times 10^2 \left(\frac{1+z}{\delta \nu_{\mathrm{GHz}} B^3}\right)^{1/2} \, \mathrm{days}
 \end{equation}
  Using the observed $\tau_{\mathrm{var}}$ and $\delta_o$, we find $B \le 3.8$ G. 
However, using $\delta=10$ \citep{1996Ap&SS.240..195B,2011MNRAS.412.1389N,2011ApJ...729...26M}, we find $R_s<19.5 \times 10^{14}$ cm and $B \sim 1.8$ G. A Doppler factor of 
17 has been reported by \cite{2014ApJ...791...53A} based on fits to monitoring
observations of OJ 287 in the radio band. Using $\delta=17$ we obtain $B \sim 1.5$ G, which is close to the value of 0.93 G found in the OJ 287 by \citet{1996Ap&SS.240..195B}.

Analysis of the long term variation of $PA$ based on our limited polarimetric
observations give a rotation rate of 5.8 degrees/day, similar to the value
of 4.92 degree/day found by \cite{2002A&A...381..408E}. This is also in
general agreement with the recent results obtained from dedicated 
optical polarimetric monitoring of blazars, which indicates that the 
rotation of the plane of optical polarization is a characteristics 
property of blazars \citep{2016MNRAS.462.1775B}. The same set of polarimetric
observations also find differences in the polarization properties of 
different sub-classes of blazars \citep{2016MNRAS.463.3365A}.
In our polarimetric observations, shown in Figure \ref{Fig:pol_freq}, for most 
of the epochs we find $P$ to decrease with wavelength. 
This is similar to that noted by  \citet{1994A&AS..107..497T}, however, 
inconsistent with that observed by  
\citet{2001A&A...376...51T}. 
 The observed WDP can be explained in terms of a two-component model identified with the jet that gives rise to the constant polarized component and the shock that gives rise to the more polarized component \citep{1991AJ....101...78V}. The presence of this stable polarized component is also evident in the position of the average Q and U values that deviate from zero in the Q vs U plane as seen 
in Figure \ref{Fig:rate}. 

If the accretion disk/host galaxy contributes significantly to the optical 
emission (in addition to the synchrotron jet emission) of OJ 287, one
 would have expected higher polarization at longer wavelengths \citep{1982ApJ...254...22M,1986ApJ...305..484S}. This is not 
 observed in any of our data except that obtained on 12 February 2016 during 
which epoch the source was in an intermediate brightness state. Also, 
 in the broadband SED of OJ 287, emission from the accretion disk is not 
 prominent \citep{2003A&A...399...33M}. Moreover, OJ 287 is a highly core dominated object\footnote{The ratio of core to extended emission is $>$995 \citep{1985ApJ...294..158A}.} and thus, the
 contribution of accretion disk to the optical emission of OJ 287 is 
 insignificant. Alternatively, in the binary black hole model of OJ 287,
thermal flares are expected when the secondary black hole crosses
the accretion disk of the primary black hole. Observations do 
indicate that such outbursts are not accompanied by increased
optical polarization. However, secondary outbursts after the major one
do show a correlated behavior in polarization as well, which could
be due to the jet of the secondary black hole getting activated. Our
polarization observations reported here are during February - April 2016, much 
later than the thermal outburst of December 2015 \citep{2016ApJ...819L..37V}. 
This along with other observational evidences outlined above indicate that 
the polarization emission 
during our observations of OJ 287 is mainly due to synchrotron processes 
happening in the jet of the source. 

\begin{figure*}
\centering
\resizebox{10cm}{7cm}{\includegraphics{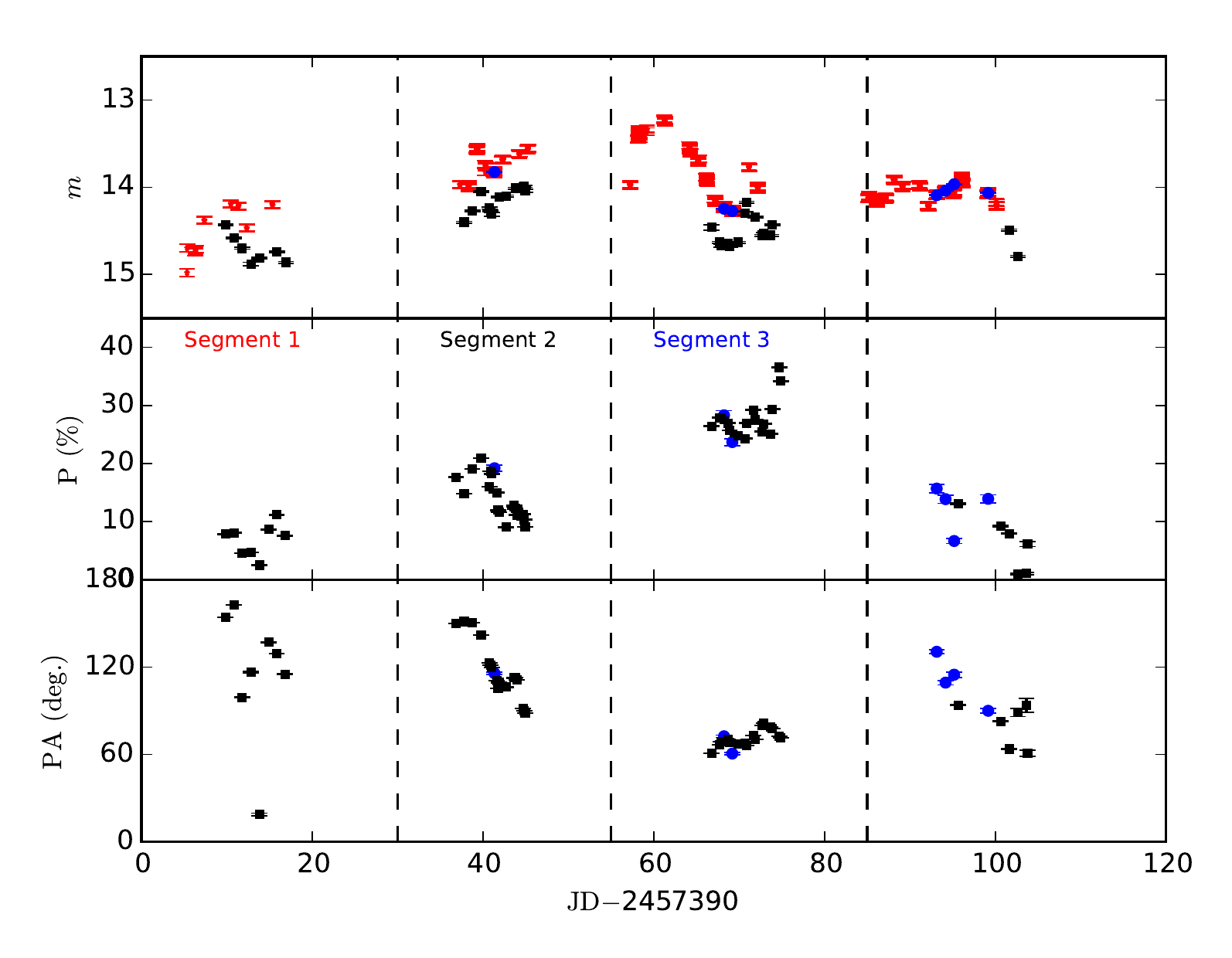}}
\caption{Top: Long term variation of $R$-band magnitude ($m_R$) as obtained by us (red) and $V$-band magnitude ($m_V$) as obtained by Steward observatory (black). The $m_R$ corresponding to our polarization measurement is shown in blue. Middle: polarization degree from Steward (black) and our (blue). Bottom: Polarization position from Steward (black) and our (blue). The data is divided into different segments for detailed analysis. A clear anticorrelation between brightness and polarization can be seen in Segment 2.}\label{Fig:pol_Steward}
\end{figure*}

 The LTPV observations  
show a clear anticorrelation between $P$ and optical brightness
as well as  between $PA$ and $P$. These results agree with  the
polarization monitoring of OJ 287 by \citet{1991AJ....101.2017S}. However, 
\citet{2009ApJ...697..985D} noticed a positive correlation between 
polarization and brightness which is contrary to what we have found. To check 
for the robustness of our results we looked for the availability of photometric and 
polarimetric data during the period of our observation. From the 
{\it Fermi} monitoring program of Steward Observatory 
\citep{2009arXiv0912.3621S} supporting the {\it Fermi} mission all-sky
survey, we 
collected 48 epochs of polarimetric and 34 epochs of $V$-band photometric data 
between the period 12 January 2016 and 15 April 2016. The data set along with 
our observations are shown in Figure \ref{Fig:pol_Steward}. The data set is 
divided into four segments based on the seasonal gaps (as shown by dotted 
lines) for detailed correlation analysis between flux and polarization 
variations. 
 
\begin{figure*}
\centering
\resizebox{15cm}{8.0cm}{\includegraphics{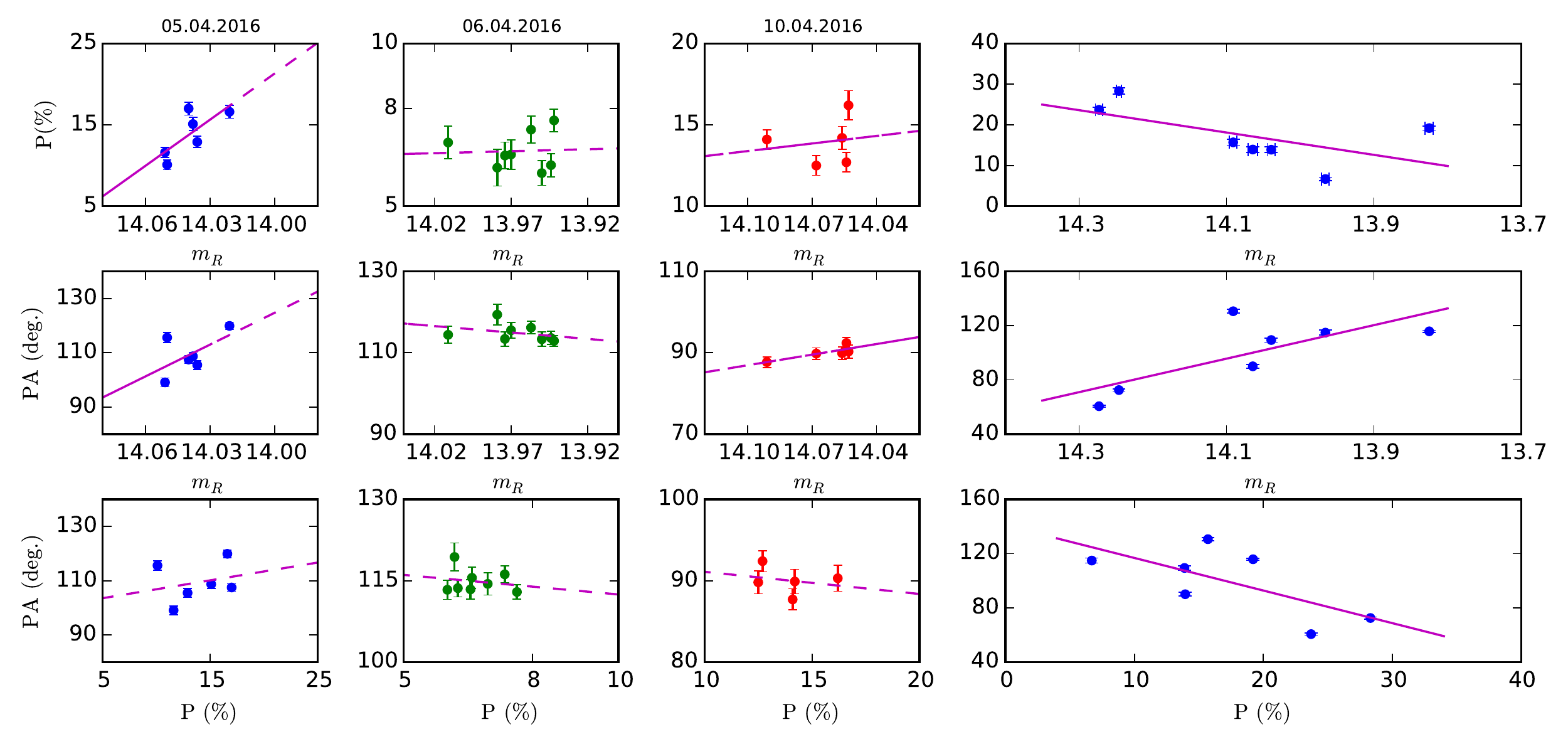}}
\caption{Plots $P$ vs $m_R$, $PA$ vs $m_R$ and $PA$ vs $P$ for INPV. The dates are indicated on the top of each panel. The right most
panel shows the correlation between different physical quantities based on LTPV observations.  The lines are the linear least squares 
fit to the data.}\label{Fig:intra_pol_rmag}.
\end{figure*}

\begin{figure*}
\centering
\resizebox{6cm}{12cm}{\includegraphics{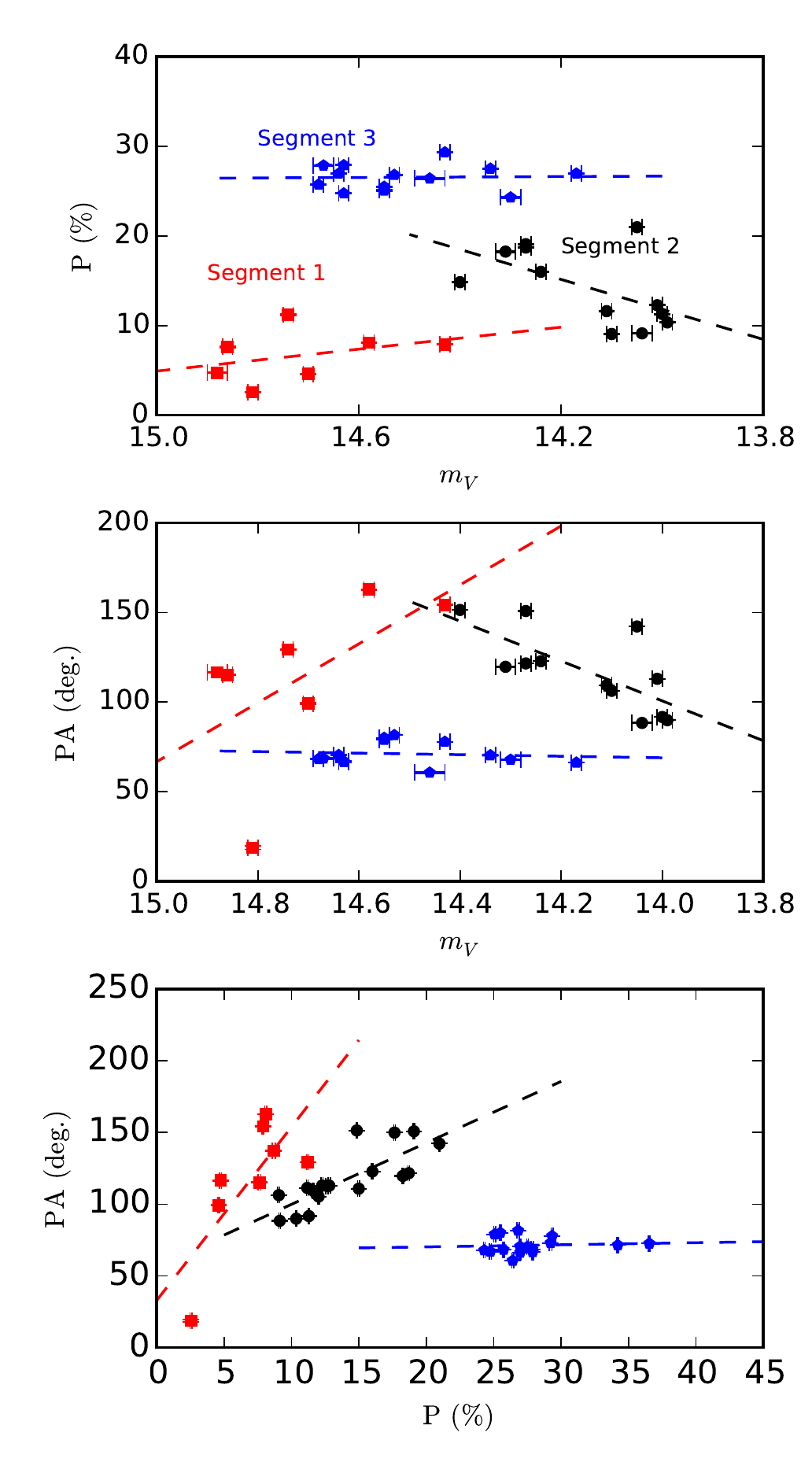}}
\caption{Polarization degree vs $V$-band magnitude (top), position angle vs $V$-band magnitude (middle) and position angle vs polarization degree (bottom) for three different segments as presented in Figure \ref{Fig:pol_Steward}. The dotted lines represent the linear fit to the data of each segment.}\label{Fig:corr_Steward}
\end{figure*}

  In Figure \ref{Fig:corr_Steward}, we show the observed relation between flux 
and polarization behavior of the source for the first three segments. The 
correlation between these quantities in segment 4 is shown in Figure 
\ref{Fig:intra_pol_rmag} as the Steward observations have only two epochs of 
data in this segment. From Figures \ref{Fig:corr_Steward} and 
\ref{Fig:intra_pol_rmag} it is evident that 
the brightness of the source positively correlates with polarization during 
segment 1 (January 2016), correlates negatively during segment 2 (February 
2016) and segment 4 (April 2016) and not show any trend during 
segment 3 (March 2016). The $PA$ positively correlates with $P$ in segments 1 and 2, however, correlates negatively in segment 4 and no correlation in segment 3. The results of the correlation analysis are given in Table \ref{Table:Pol_corr_Steward}.
 
In the shock-in-jet model of blazar variability, a positive correlation 
between flux and polarization variations is expected \citep{1985ApJ...298..114M} which could be due to the magnetic field getting aligned because of the shock. Alternatively, if the flux variability is due to the emergence of a new blob of plasma (identified as a VLBI scale knot) which has either a chaotic magnetic field or a magnetic field that is misaligned with the large scale field, an anticorrelation between flux and polarization variations can be expected \citep{2002A&A...385...55H,2002ApJ...568...99H}. The observed correlation and anticorrelation between total flux and polarized flux can also be explained by changes in the trajectories of the shocks propagating down the relativistic jets as postulated in the ``swinging jets'' model of \citet{1992A&A...259..109G}. From the observations of OJ 287 reported here we find varied behavior between flux and polarization variations, which could happen because of the presence of more than one emission region in the jet of OJ 287 \citep{2008Natur.452..966M} or due to the 
interaction between the jet and accretion disk \citep{2008Natur.452..851V,2016ApJ...819L..37V}.
Near simultaneous flux and polarization observations of blazars are very 
limited and observations on a large sample of blazars are needed, which will give important leads to our understanding on the emission processes in blazars.
 
 \section{Conclusions}\label{sec:conclusion}
We have carried out photometric (40 nights) and polarimetric (9 epochs) 
observations of OJ 287 coinciding with its high brightness state 
during December 2015 $-$ February 2016. 
The key findings are summarized below:
 \begin{enumerate}
 \item From 21 nights of INOV observations we found
 the source to show INOV on few nights. Using $C$-statistics
 we found the $DC$ of INOV as 30 \%, which increases to 45\% and 71 \% on using
 the $F$-statistics and $\chi^2$-statistics respectively. On nights when INOV 
is observed, $\Psi$ is larger than 3\%. The observed large amplitude ($>$ 3\%) and high $DC$ of 
 INOV are similar to that known for blazars.
 \item We find the shortest flux variability time scale of $142\pm 38$ min on April 07, 2016. Using this
 we put constraints on the size of the emitting region and magnetic
 field strength as $2.28\times 10^{14}$ ($19.5 \times 10^{14}$) cm and 3.8 (1.8) G using $\delta=1.17$ (10) respectively.
 \item Considering LTOV, we find
 that OJ 287 has varied by about 2 mag during the period of our 
 observations. During this period, it showed a maximum and 
 minimum brightness of 13.20 $\pm$ 0.04 mag and 14.98 $\pm$ 0.04 mag 
respectively.  A change of $\sim$1 mag was noticed in March within
10 days. 
 \item From polarimetric observations, we find OJ 287 showed both 
 INPV and LTPV. Considering the polarization
 variations during February to April 2016, minimum and maximum $P$ of 
 6 $\pm$ 0.3 \% and 28.3 $\pm$ 0.8 \% in $R$-band was observed. During the same period $PA$ varied between 60.6 $\pm$ 0.8 degrees  and 130.6 $\pm$ 1.3 degrees respectively. 
 \item In U v/s Q plane, the 
 average Q and U deviate from zero, indicating the presence of two
 optically thin synchrotron emission components contributing to the 
 polarized emission from OJ 287 jet.
 \item The $P$ in different wavebands are correlated, with
 the polarization at shorter wavelengths generally larger than 
 at longer wavelengths, thus showing a wavelength dependent polarization
 behavior. This demands that the observed polarization is due to synchrotron process happening in the jet of the source.
\item During most of the observing period an anticorrelation is observed 
between flux and polarization variations. A wide variety of correlations are 
also noticed between $PA$ and $P$ as well as between $PA$ and brightness. 
Such a variety of relations observed between flux and polarization variations 
might be because of the presence of more than one emission components in the 
jet of OJ 287.
\end{enumerate} 

\vspace*{0.05cm}

\acknowledgments
We are grateful for the comments and suggestions by the anonymous referee, which helped to improve the manuscript. It is our pleasure to thank Professor A. V. Raveendran and Mr. G. Srinivasulu for their valuable suggestions and timely help for the efficient operation of the photo-polarimeter. We also thank K. Sagayanathan, A. K. Venkataramana, R. Baskar, S. Surendharnath, A. Muniyandi, A. Ramachandran, M. Muniraj and the personnel of the technical divisions for their support to carry out 
the observations presented in this paper. 
Data from the Steward Observatory spectropolarimetric monitoring project were used. This program is supported by Fermi Guest Investigator grants NNX08AW56G, NNX09AU10G, NNX12AO93G, and NNX15AU81G.

 \bibliographystyle{apj}
 \bibliography{ref}
 
 \begin{table*}
 \caption{Log of photometric observation. Column information are as follows: (1) date of observations; (2) number of data points in DLC; (3) exposure time in second; (4) duration of monitoring in hour. This Table is published in its entirety in the machine-readable format. A portion is shown here for guidance regarding its form
and content.}
 \begin{minipage}{1.0\textwidth}
 	\centering {
 	\small\addtolength{\tabcolsep}{5pt}
 	\begin{tabular}{ r r r r}\hline 
 	Date & $N$ & Exp. time & Duration \\
 	 (dd.mm.yyyy)    &   &  (s)      & (h) \\
 	 (1) & (2) & (3)     & (4) \\ \hline    
 	07.01.2016	 &	 2	 &	 300	 &	 0.2	\\
 	08.01.2016	 &	 3	 &	 900	 &	 0.9	\\\hline 
 	\label{Table:log}
	\end{tabular} }
	 	\end{minipage}
	 \end{table*}

  \begin{table*}
  \caption{Results of polarization observations. Column details are as follows: (1) date of observation; (2) observing band ($R'$ is the integrated polarization in the $VRI$ spectral region); (3) time in Julian Day; (4) degree of polarization in per cent; (5) error in degree of polarization in per cent; (6) polarization position angle in degree; (7) error in position angle in degree. 
  }
  \begin{minipage}{1.0\textwidth}
  	\centering {
  	\small\addtolength{\tabcolsep}{6pt}
  	\begin{tabular}{ r r r r r r r}\hline 
  	Date     & Band  & JD   &  $P$    & $P_\mathrm{error}$  & $PA$       & $PA_\mathrm{error}$ \\
 (dd.mm.yyyy) &       &      & (\%)  &  (\%)    & (deg.)   & (deg.) \\
  	 (1)     & (2)   & (3)  & (4)	& (5)      & (6)      & (7)      \\ \hline    
 12.02.2016  &  $U$ & 2457431.3201 & 16.3 & 1.1 & 110.9 & 1.9 \\
 12.02.2016  &  $B$ & 2457431.3201 & 18.8 & 0.8 & 118.0 & 1.2 \\
 12.02.2016  &  $V$ & 2457431.3648 & 19.8 & 1.0 & 119.0 & 1.4 \\
 12.02.2016  &  $R$ & 2457431.3372 & 19.2 & 0.5 & 115.8 & 0.8 \\
 12.02.2016  &  $I$ & 2457431.3104 & 16.6 & 0.3 & 118.0 & 1.0 \\
 12.02.2016  &  $R'$ & 2457431.2796 & 18.9 & 0.3 & 116.9 & 0.5 \\
 08.03.2016  &  $U$ & 2457456.2399 & 33.0 & 1.8 &  62.4 & 1.7 \\
 08.03.2016  &  $B$ & 2457456.2399 & 28.6 & 1.5 &  59.9 & 1.6 \\
 08.03.2016  &  $R'$ & 2457456.1932 & 22.6 & 0.2 &  60.8 & 0.3 \\
 09.03.2016  &  $R'$ & 2457457.1786 & 27.7 & 0.5 &  61.2 & 0.5 \\
 10.03.2016  &  $U$ & 2457458.2526 & 38.7 & 1.4 &  74.3 & 1.1 \\
 10.03.2016  &  $B$ & 2457458.2526 & 32.1 & 1.1 &  64.9 & 1.1 \\
 10.03.2016  &  $V$ & 2457458.2475 & 27.8 & 0.8 &  71.0 & 0.9 \\
 10.03.2016  &  $R$ & 2457458.2230 & 28.3 & 0.8 &  72.5 & 0.8 \\
 10.03.2016  &  $I$ & 2457458.2870 & 22.5 & 0.9 &  67.3 & 1.2 \\
 10.03.2016  &  $R'$ & 2457458.2012 & 27.4 & 0.4 &  71.6 & 0.4 \\
 11.03.2016  &  $U$ & 2457459.2242 & 32.2 & 3.0 &  63.2 & 3.1 \\
 11.03.2016  &  $B$ & 2457459.2242 & 27.2 & 1.6 &  58.6 & 1.8 \\
 11.03.2016  &  $V$ & 2457459.2232 & 30.0 & 0.9 &  63.5 & 0.9 \\
 11.03.2016  &  $R$ & 2457459.1924 & 23.7 & 0.6 &  60.6 & 0.8 \\
 11.03.2016  &  $R'$ & 2457459.1703 & 26.7 & 0.4 &  61.1 & 0.5 \\
 04.04.2016  &  $U$ & 2457483.1590 & 24.2 & 1.9 & 114.3 & 2.3 \\
 04.04.2016  &  $B$ & 2457483.1590 & 15.6 & 1.1 & 126.5 & 2.1 \\
 04.04.2016  &  $V$ & 2457483.1690 & 12.7 & 0.7 & 123.9 & 1.7 \\
 04.04.2016  &  $R$ & 2457483.1331 & 15.7 & 0.7 & 130.6 & 1.3 \\
 04.04.2016  &  $I$ & 2457483.2047 & 12.3 & 0.7 & 121.5 & 1.7 \\
 04.04.2016  &  $R'$ & 2457483.1112 & 16.8 & 0.4 & 127.5 & 0.6 \\
 05.04.2016  &  $R$ & 2457484.1184 & 17.0 & 0.8 & 107.5 & 1.4 \\
 05.04.2016  &  $R$ & 2457484.1451 & 15.1 & 0.8 & 108.7 & 1.5 \\
 05.04.2016  &  $R$ & 2457484.1666 & 11.6 & 0.6 &  99.1 & 1.6 \\
 05.04.2016  &  $R$ & 2457484.1821 & 10.1 & 0.6 & 115.6 & 1.8 \\
 05.04.2016  &  $R$ & 2457484.1994 & 12.9 & 0.7 & 105.5 & 1.6 \\
 05.04.2016  &  $R$ & 2457484.2221 & 16.6 & 0.8 & 119.9 & 1.4 \\
 06.04.2016  &  $R$ & 2457485.1145 & 7.6  & 0.3 & 112.9 & 1.3 \\
 06.04.2016  &  $R$ & 2457485.1320 & 6.5  & 0.4 & 113.4 & 1.7 \\
 06.04.2016  &  $R$ & 2457485.1491 & 6.5  & 0.4 & 115.5 & 1.9 \\
 06.04.2016  &  $R$ & 2457485.1666 & 7.3  & 0.4 & 116.1 & 1.5 \\
 06.04.2016  &  $R$ & 2457485.1841 & 6.1  & 0.5 & 119.3 & 2.5 \\
 06.04.2016  &  $R$ & 2457485.2012 & 6.9  & 0.5 & 114.4 & 2.0 \\
 06.04.2016  &  $R$ & 2457485.2187 & 6.2  & 0.3 & 113.6 & 1.6 \\
 06.04.2016  &  $R$ & 2457485.2362 & 6.0  & 0.3 & 113.3 & 1.7 \\
 10.04.2016  &  $R$ & 2457489.1125 & 14.1 & 0.6 &  87.7 & 1.3 \\
 10.04.2016  &  $R$ & 2457489.1398 & 12.5 & 0.6 &  89.8 & 1.4 \\
 10.04.2016  &  $R$ & 2457489.1634 & 12.7 & 0.6 &  92.4 & 1.3 \\
 10.04.2016  &  $R$ & 2457489.1869 & 14.2 & 0.7 &  89.9 & 1.5 \\
 10.04.2016  &  $R$ & 2457489.2115 & 16.2 & 0.9 &  90.3 & 1.6 \\ \hline

 	\end{tabular} }
 	 	\end{minipage}
 	 	\label{Table:pol-data}
 	 \end{table*}
 
 \begin{table*}
 \caption{Intra-night variability properties. Column information are: (1) date of observation; (2) and (3) INOV amplitudes in \%; (4) and (5) $F$-values computed for the OJ 287 DLCs relative to the steadiest pair of comparison stars (star 4 and star 10) on any night; (6) variability status according to $F$-statistics; (7) and (8) values of $C$ for the two OJ 287 DLCs relative to the two comparison stars (star 4 and star 10); (9) variability status as per $C$-statistics; (10) $\chi^2$ value; (11) critical value $\chi^2_{\alpha=0.01, \nu}$; (12) variability status as per $\chi^2$-statistics; (13) time duration of observation in hour.}
 \begin{minipage}{1.0\textwidth}
 	\centering {
 	\small\addtolength{\tabcolsep}{0.5pt}
     \begin{tabular}{ r r r r r r r r r r r r r}\hline   
    Date  & $\Psi$1 & $\Psi$2 & $F_1$ & $F_2$ & Status & $C_1$ & $C_2$ & Status & $\chi^2$ & $\chi^2_{\alpha=0.01, \nu}$ & Status & $dt$ (hrs.) \\  
    (dd.mm.yyyy) & ( \% )   &   (\%)       &       &  & & & &  & & & & \\
    (1)  & (2) & (3) & (4) & (5) & (6) & (7) & (8) & (9) & (10) & (11) & (12) & (13)\\ \hline

10.02.2016	 &	 4.70	 &	 4.00	 &	  1.380	 &	 1.171	 &	 NV	 &	 1.175	 &	 1.082	 &	 NV	 &	 163.208	 &	 42.980	 &	 V	 &	 2.2	 \\
12.02.2016	 &	 4.00	 &	 5.80	 &	  1.026	 &	 1.365	 &	 NV	 &	 1.013	 &	 1.168	 &	 NV	 &	 122.299	 &	 46.963	 &	 V	 &	 3.1	 \\
13.02.2016	 &	 2.20	 &	 2.70	 &	  0.598	 &	 1.626	 &	 NV	 &	 0.773	 &	 1.275	 &	 NV	 &	 26.847	 &	 37.566	 &	 NV	 &	 2.5	 \\
29.02.2016	 &	 11.80	 &	 11.10	 &	  12.912	 &	 10.753	 &	 V	 &	 3.593	 &	 3.279	 &	 V	 &	 1932.985	 &	 48.278	 &	 V	 &	 2.9	 \\
03.03.2016	 &	 4.70	 &	 2.10	 &	  1.325	 &	 0.308	 &	 NV	 &	 1.151	 &	 0.555	 &	 NV	 &	 281.167	 &	 38.932	 &	 V	 &	 2.2	 \\
06.03.2016	 &	 9.00	 &	 10.30	 &	  17.322	 &	 14.964	 &	 V	 &	 4.162	 &	 3.868	 &	 V	 &	 1020.255	 &	 37.566	 &	 V	 &	 2.2	 \\
07.03.2016	 &	 4.90	 &	 5.50	 &	  6.707	 &	 9.300	 &	 V	 &	 2.590	 &	 3.050	 &	 V	 &	 284.627	 &	 34.805	 &	 V	 &	 1.8	 \\
08.03.2016	 &	 7.20	 &	 7.20	 &	  42.775	 &	 43.884	 &	 V	 &	 6.540	 &	 6.625	 &	 V	 &	 490.736	 &	 41.638	 &	 V	 &	 2.0	 \\
09.03.2016	 &	 4.80	 &	 4.80	 &	  9.294	 &	 7.986	 &	 V	 &	 3.049	 &	 2.826	 &	 V	 &	 218.527	 &	 37.566	 &	 V	 &	 1.8	 \\
10.03.2016	 &	 3.50	 &	 4.10	 &	  2.440	 &	 3.003	 &	 NV	 &	 1.562	 &	 1.733	 &	 NV	 &	 99.288	 &	 34.805	 &	 V	 &	 2.3	 \\
11.03.2016	 &	 3.70	 &	 3.90	 &	  3.916	 &	 4.508	 &	 V	 &	 1.979	 &	 2.123	 &	 NV	 &	 132.167	 &	 46.963	 &	 V	 &	 2.8	 \\
27.03.2016	 &	 3.89	 &	 2.38	 &	  1.310	 &	 0.471	 &	 NV	 &	 1.145	 &	 0.686	 &	 NV	 &	 29.150	 &	 38.932	 &	 NV	 &	 2.1	 \\
29.03.2016	 &	 3.30	 &	 3.00	 &	  4.958	 &	 4.507	 &	 V	 &	 2.227	 &	 2.123	 &	 NV	 &	 57.634	 &	 36.191	 &	 V	 &	 1.8	 \\
30.03.2016	 &	 2.70	 &	 3.60	 &	  0.747	 &	 1.174	 &	 NV	 &	 0.864	 &	 1.084	 &	 NV	 &	 47.865	 &	 45.642	 &	 V	 &	 2.5	 \\
31.03.2016	 &	 2.00	 &	 2.70	 &	  1.420	 &	 2.111	 &	 NV	 &	 1.192	 &	 1.453	 &	 NV	 &	 27.399	 &	 40.289	 &	 NV	 &	 2.3	 \\
03.04.2016	 &	 2.20	 &	 2.90	 &	  1.476	 &	 2.342	 &	 NV	 &	 1.215	 &	 1.530	 &	 NV	 &	 23.993	 &	 38.932	 &	 NV	 &	 2.0	 \\
04.04.2016	 &	 2.40	 &	 2.40	 &	  1.027	 &	 1.262	 &	 NV	 &	 1.013	 &	 1.123	 &	 NV	 &	 24.070	 &	 36.191	 &	 NV	 &	 2.0	 \\
05.04.2016	 &	 3.00	 &	 3.20	 &	  0.987	 &	 1.305	 &	 NV	 &	 0.993	 &	 1.142	 &	 NV	 &	 37.446	 &	 42.980	 &	 NV	 &	 2.2	 \\
06.04.2016	 &	 4.20	 &	 5.80	 &	  1.200	 &	 2.071	 &	 NV	 &	 1.095	 &	 1.439	 &	 NV	 &	 69.109	 &	 42.980	 &	 V	 &	 2.1	 \\
07.04.2016	 &	 10.10	 &	 10.40	 &	  14.908	 &	 13.025	 &	 V	 &	 3.861	 &	 3.609	 &	 V	 &	 484.825	 &	 42.980	 &	 V	 &	 2.2	 \\
10.04.2016	 &	 4.50	 &	 4.70	 &	  4.238	 &	 6.485	 &	 V	 &	 2.059	 &	 2.547	 &	 NV	 &	 110.030	 &	 40.289	 &	 V	 &	 2.0	 \\ \hline
\end{tabular} }
 	\label{Table:INOV_prop}
 	\end{minipage}
 \end{table*}

 \begin{table*}
 \caption{Results on the correlation analysis between photometric and
polarimetric observations. Columns are listed as follows: (1) date of observation; (2) correlation between datasets; (3) Pearson correlation coefficient ($r_p$); (4) $p$ value for no correlation.}
\begin{minipage}{1.0\textwidth}
        \centering {
        \small\addtolength{\tabcolsep}{7pt}
 	\begin{tabular}{ r r r r }\hline 
 	Date     & Parameter       & $r_p$   &  Significance   \\
 	  (1)    & (2)             & (3)  & (4) \\ \hline
        All      &  $F - P$         & -0.66    & 0.000               \\
                 &  $F - PA$  & +0.75    & 0.000                \\
                 & $P - PA$  & -0.62    & 0.001               \\ \hline
      05.04.2016 & $F - P$         & +0.73    & 0.093                \\
                 & $F - PA$  & +0.57    & 0.230                \\
                 & $P - PA$  & +0.24    & 0.635               \\ \hline
      06.04.2016 & $F - P$         & +0.04    & 0.909                \\
                 &  $F - PA$  & -0.33    & 0.416                \\
                 &  $P - PA$  & -0.19    & 0.646               \\ \hline
      10.04.2016 & $F - P$         & +0.16    & 0.787                \\
                 &  $F - PA$  & +0.83    & 0.079               \\
                 & $P - PA$  & -0.24    & 0.694               \\ \hline
	\end{tabular} }
	 	\end{minipage}
	 	\label{Table:Pol_prop}
	 \end{table*}

 \begin{table*}
 \caption{Intra-night polarization properties. Columns are listed as follows: (1) date of observation; (2) and (3) mean and standard deviation of degree of polarization;  (4) Fractional polarization variability; (5) $\chi^2$ of $P$; (6) Variable (V) /Non-variable (NV); (7) to (11) are the same as (2) to (6) but for position angle.}
 \begin{minipage}{0.9\textwidth}
 	\centering {
 	\small\addtolength{\tabcolsep}{4pt}
     \begin{tabular}{ r r r r r r r r r r r}\hline
     Date  & $<P>$ & $\sigma_{P}$ & F.V. & $\chi^2_{P}$ & Status & $<PA>$ & $\sigma_{PA}$ & F.V. &  $\chi^2_{PA}$ & Status \\  
     (dd.mm.yyyy) & ( \% )   &  &      &       &  & ( $\degree$ ) &  & &  & \\
      (1)  & (2) & (3) & (4) & (5) & (6) & (7) & (8) & (9) & (10) & (11)\\ \hline
     
05.04.2016 &	 13.24 &	 2.552 &	 0.255 &	 80.230 &	 V  &	 109.57	 & 6.759 &	 0.095 &	 117.480 &	 V \\
06.04.2016 &	 6.73  &	 0.539 &	 0.119 &	 15.701 &	 NV  &	 114.39	 & 2.015 &	 0.028 &	 7.478	 & 	 NV \\
10.04.2016 &	 13.62 &	 1.328 &	 0.129 &	 15.380 &	 V  &	 90.01	 & 1.496 &	 0.026 &	 6.598	 & 	 NV \\ \hline
	\end{tabular} }
 	\label{Table:INPV_prop}
 	\end{minipage}
 \end{table*}

\begin{table*}
\caption{ Correlation analysis between flux and polarization of OJ 287. Columns are listed as follows: (1) Segment number; (2) Pearson correlation coefficient of $P$ vs $V$-band magnitude; (3) $PA$ vs $V$; (4) $P$ vs $PA$. The $p$ value of no correlation is written 
within brackets. }
\begin{minipage}{1.0\textwidth}
        \centering {
        \small\addtolength{\tabcolsep}{7pt}
 	\begin{tabular}{ r r r r }\\ \hline 
     Segment   & $P$ vs $V$ & $PA$ vs $V$ & $PA$ vs $P$ \\
     (1)    & (2)              & (3)  & (4) \\ \hline
     Segment 1 & -0.34 (0.45)  & -0.56 (0.19) & 0.74 (0.0327) \\
     Segment 2 &  0.55 (0.05)  & 0.70 (0.01)  & 0.78 (0.0001) \\
     Segment 3 &  -0.02 (0.92) & 0.10 (0.72)  & 0.08 (0.7566)  \\ \hline
	\end{tabular} }
\end{minipage}\label{Table:Pol_corr_Steward}
\end{table*}
\end{document}